\documentclass[12pt]{article}
\usepackage{epsfig}

\newcommand{\cd}{\makebox[0.08cm]{$\cdot$}}
\newcommand{\MSbar} {\hbox{$\overline{\hbox{\tiny MS}}$}}

\newcommand{\ket}[1]{\,\left|\,{#1}\right\rangle}
\begin {document}
\begin{flushright}
{\small
SLAC--PUB--10812\\
October 2004\\}
\end{flushright}

\vfill

\begin{center}
{{\bf\LARGE New Results in Light-Front\\[1ex]
Phenomenology}\footnote{Work supported by Department of Energy
contract DE--AC02--76SF00515.}}

\bigskip
{\it Stanley J. Brodsky \\
Stanford Linear Accelerator Center \\
Stanford University, Stanford, California 94309 \\
E-mail:  sjbth@slac.stanford.edu}
\medskip
\end{center}

\vfill

\begin{center}
{\it Presented at \\
LightCone 2004 \\
Amsterdam, The Netherlands \\
16--20 August 2004}\\
\end{center}

\vfill \newpage

\vfill

\begin{center}
{\bf\large Abstract }
\end{center}

The light-front quantization of gauge theories in light-cone
gauge provides a frame-independent wavefunction representation of
relativistic bound states, simple forms for current matrix elements,
explicit unitarity, and a trivial vacuum.   In this talk I review
the theoretical methods and constraints which can be used to
determine these central elements of QCD phenomenology.  The freedom
to choose the light-like quantization four-vector provides an
explicitly covariant formulation of light-front quantization and can
be used to determine the analytic structure of light-front wave
functions and define a kinematical definition of angular momentum.
The AdS/CFT correspondence of large $N_C$ supergravity theory in
higher-dimensional  anti-de Sitter space with supersymmetric QCD in
4-dimensional space-time  has interesting implications for hadron
phenomenology in the conformal limit, including an all-orders
demonstration of counting rules for exclusive processes.
String/gauge duality also predicts the QCD power-law behavior of
light-front Fock-state hadronic wavefunctions with arbitrary orbital
angular momentum at high momentum transfer.  The form of these
near-conformal wavefunctions can be used as an initial ansatz for a
variational treatment of the light-front QCD Hamiltonian.  The
light-front Fock-state wavefunctions encode the bound state
properties of hadrons in terms of their quark and gluon degrees of
freedom at the amplitude level.  The nonperturbative Fock state
wavefunctions contain intrinsic gluons, and sea quarks at any scale
$Q$ with asymmetries such as $ s(x) \ne \bar s(x)$, $\bar u(x) \ne
\bar d(x).$  Intrinsic charm and bottom quarks appear at large $x$
in the light-front wavefunctions since this minimizes the invariant
mass and off-shellness of the higher Fock state.  In the case of
nuclei, the Fock state expansion contains ``hidden color" states
which cannot be classified in terms of nucleonic degrees of freedom.
I also briefly review recent analyses which shows that some
leading-twist phenomena such as the diffractive component of deep
inelastic scattering, single-spin asymmetries, nuclear shadowing and
antishadowing cannot be computed from the LFWFs of hadrons in
isolation.

\vfill
\newpage

\section{Introduction}

A central problem in nonperturbative quantum chromodynamics is to
determine not only the masses but also the wavefunctions of hadronic
bound states.  Relativity and quantum mechanics requires that a
hadron fluctuates not only in coordinate space, spin, and color, but
also in the number of quanta.  The light-front Hamiltonian
formulation of quantum chromodynamics provides a comprehensive
formulation for determining not only the spectrum of the theory, but
also the complete set of light-front Fock state wavefunctions
$\psi_{n/H}(x_i,\vec k_{\perp i},\lambda_i)$ which encode the bound
state properties  of hadrons in terms of their fundamental quark and
gluon degrees of freedom at the amplitude level.

Formally, the light-front expansion is constructed by quantizing QCD
at fixed light-cone time \cite{Dirac:1949cp} $\tau = t + z/c$ and
forming the invariant light-front Hamiltonian: $ H^{QCD}_{LF} = P^+
P^- - {\vec P_\perp}^2$ where $P^\pm = P^0 \pm
P^z$~\cite{Brodsky:1997de}.  The momentum generators $P^+$ and $\vec
P_\perp$ are kinematical; {\em i.e.}, they are independent of the
interactions. The generator $P^- = i {d\over d\tau}$ generates
light-cone time translations, and the eigen-spectrum of the Lorentz
scalar $ H^{QCD}_{LF}$ gives the mass spectrum of the color-singlet
hadron states in QCD together with their respective light-front
wavefunctions.  For example, the proton state satisfies: $
H^{QCD}_{LF} \ket{\psi_p} = M^2_p \ket{\psi_p}$.

The light-front (LF) quantization of QCD in light-cone gauge $A^+=0$
has a number of remarkable advantages, including explicit unitarity,
a physical Fock expansion, the absence of ghost degrees of freedom,
and the decoupling properties needed to prove factorization theorems
in high momentum transfer inclusive and exclusive reactions.  Prem
Srivastava and I have given  a systematic
derivation~\cite{Srivastava:2000cf} of LF-quantized gauge theory
using the Dirac method of constraints.  The free theory gauge field
is shown to satisfy the Lorentz condition as an operator equation as
well as the light-cone gauge condition.  Its propagator is found to
be transverse with respect to both its four-momentum and the gauge
direction.  The interaction Hamiltonian of QCD has a form resembling
that of covariant theory, except for additional instantaneous
interactions which can be treated systematically.  The QCD $\beta$
function computed in the light-cone gauge agrees with that known in
the conventional framework.  In the case of the electroweak theory,
spontaneous symmetry breaking is realized in LF quantization by the
appearance of zero modes of the Higgs field.  Light-front
quantization leads to an elegant ghost-free theory of massive gauge
particles, automatically incorporating the Lorentz and 't Hooft
conditions, as well as the Goldstone boson equivalence
theorem~\cite{Srivastava:2002mw}.

The expansion of the proton eigensolution $\ket{\psi_p}$ on the
color-singlet $B = 1$, $Q = 1$ eigenstates $\{\ket{n} \}$ of the
free Hamiltonian $ H^{QCD}_{LF}(g = 0)$ gives the light-front Fock
expansion:
\begin{eqnarray}
\ket{ \psi_p(P^+, {\vec P_\perp} )} &=& \sum_{n}\ \prod_{i=1}^{n} {{\rm
d}x_i\, {\rm d}^2
{\vec k_{\perp i}} \over \sqrt{x_i}\, 16\pi^3} \,  \ 16\pi^3  \
\delta\left(1-\sum_{i=1}^{n} x_i\right)\, \delta^{(2)}\left(\sum_{i=1}^{n}
{\vec k_{\perp
i}}\right) \label{a318}
\\
&& \rule{0pt}{4.5ex} \times \psi_{n/H}(x_i,{\vec k_{\perp i}},
\lambda_i) \ket{ n;\, x_i P^+, x_i {\vec P_\perp} + {\vec k_{\perp
i}}, \lambda_i}. \nonumber
\end{eqnarray}
The light-cone momentum fractions $x_i = k^+_i/P^+$ and ${\vec k_{\perp i}}$ represent
the relative momentum coordinates of the QCD constituents.  The physical transverse
momenta are ${\vec p_{\perp i}} = x_i {\vec P_\perp} + {\vec k_{\perp i}}.$ The
$\lambda_i$ label the light-cone spin projections $S^z$ of the quarks and gluons along
the quantization direction $z$. Each Fock component has the invariant mass squared
\begin{equation}\mathcal{M}^2_n = (\sum^n_{i=1} k_i^\mu)^2
= \sum^n_{i=1}{k^2_{\perp i} + m^2_i\over x_i}.
\end{equation}
The physical gluon polarization vectors $\epsilon^\mu(k,\ \lambda = \pm 1)$ are specified
in light-cone gauge by the conditions $k \cdot \epsilon = 0,\ \eta \cdot \epsilon =
\epsilon^+ = 0.$ The gluonic quanta which appear in the Fock states thus have physical
polarization $\lambda = \pm 1$ and positive metric. Since each Fock particle is on its
mass shell in a Hamiltonian framework, $k^- = k^0-k^z= {k^2_\perp + m^2\over k^+}$. One
cannot truncate the LF expansion; the expansion sum runs over all $n,$ beginning with the
lowest valence state. The probability of massive Fock states with invariant mass
$\mathcal{M}$ falls-off at least as fast as $1/\mathcal{M}^2.$

Because they are defined at fixed light-front time $\tau = t + z/c$
(Dirac's ``Front Form"), LFWFs have the remarkable property of being
independent of the hadron's four-momentum.  In contrast, in
equal-time quantization,  a Lorentz boost mixes dynamically with the
interactions, so that computing a wavefunction in a new frame at
fixed $t$ requires solving a nonperturbative problem as complicated
as the Hamiltonian eigenvalue problem itself.  The LFWFs are
properties of the hadron itself; they are thus universal and process
independent.

The light-front Fock state expansion provides important perspectives
on the quark and gluon distributions of hadrons.  For example, there
is no scale $Q_0$ where one can limit the quark content of a hadron
to valence quarks.  The nonperturbative Fock state wavefunctions
contain intrinsic gluons, strange quarks, charm quarks, etc., at any
scale.  The internal QCD interactions lead to asymmetries such as $
s(x) \ne \bar s(x)$, $\bar u(x) \ne \bar d(x)$ and intrinsic charm
and bottom distributions at large $x$ since this minimizes the
invariant mass and off-shellness of the higher Fock state.  In the
case of nuclei, the Fock state expansion contains hidden color
states which cannot be classified in terms of nucleonic degrees of
freedom.  However, some leading-twist phenomena such as the
diffractive component of deep inelastic scattering, single-spin
asymmetries, nuclear shadowing and antishadowing cannot be computed
from the LFWFs of hadrons in isolation. These issues are reviewed in
Section 5 below.

One of the important aspects of fundamental hadron structure is the
presence of non-zero orbital angular momentum in the bound-state
wave functions.  The evidence for a ``spin crisis" in the
Ellis-Jaffe sum rule signals a significant orbital contribution in
the proton wave function~\cite{Jaffe:1989jz,Ji:2002qa}.  The Pauli
form factor of nucleons is computed from the overlap of LFWFs
differing by one unit of orbital angular momentum $\Delta L_z= \pm
1$.  Thus the fact that the anomalous moment of the proton is
non-zero requires nonzero orbital angular momentum in the proton
wavefunction~\cite{BD80}.  In the light-front method, orbital
angular momentum is treated explicitly; it includes the orbital
contributions induced by relativistic effects, such as the
spin-orbit effects normally associated with the conventional Dirac
spinors.  Angular momentum conservation for each Fock state implies
\begin{equation}
J^z= \sum_i^{n} S^z_i + \sum_i^{n-1} L^z_i
\end{equation}
where $L^z_i$ is one of the $n-1$ relative orbital angular momenta.

One can also define the light-front Fock expansion using a covariant generalization of
light-front time: $\tau=x \cd \omega$. The four-vector $\omega$, with $\omega^2 = 0$,
determines the orientation of the light-front plane; the freedom to choose $\omega$
provides an explicitly covariant formulation of light-front quantization~\cite{cdkm}: all
observables such as matrix elements of local current operators, form factors, and cross
sections are light-front invariants -- they must be independent of $\omega_\mu.$ In
recent work, Dae Sung Hwang, John Hiller, Volodya Karmonov~\cite{Brodsky:2003pw}, and I
have studied the analytic structure of LFWFs using the explicitly Lorentz-invariant
formulation of the front form.  Eigensolutions of the Bethe-Salpeter equation have
specific angular momentum as specified by the Pauli-Lubanski vector.  The corresponding
LFWF for an $n$-particle Fock state evaluated at equal light-front time $\tau =
\omega\cdot x$ can be obtained by integrating the Bethe-Salpeter solutions over the
corresponding relative light-front energies.  The resulting LFWFs $\psi^I_n(x_i, k_{\perp
i})$ are functions of the light-cone momentum fractions $x_i= {k_i\cdot \omega / p \cdot
\omega}$ and the invariant mass  of the constituents $\mathcal{M}_n,$ each multiplying
spin-vector and polarization tensor invariants which can involve $\omega^\mu.$  They are
eigenstates of the Karmanov--Smirnov kinematic angular momentum
operator~\cite{ks92,cdkm}.
\begin{equation}\label{ac1}
\vec{J} = -i[\vec{k}\times
\partial/\partial\vec{k}\,]-i[\vec{n}\times
\partial/\partial\vec{n}] +\frac{1}{2}\vec{\sigma},
\end{equation}
where $\vec n$ is the spatial component of $\omega$ in the
constituent rest frame ($\vec{\mathcal{P}}=\vec 0$).  Although this
form is written specifically in the constituent rest frame, it can
be generalized to an arbitrary frame by a Lorentz boost.

Normally the generators of angular rotations in the LF formalism
contain interactions, as in the Pauli--Lubanski formulation;
however, the LF angular momentum operator can also be represented in
the kinematical form (\ref{ac1}) without interactions. The key term
is the generator of rotations of the LF plane
$-i[\vec{n}\times\partial/\partial\vec{n}]$ which replaces the
interaction term; it appears only in the explicitly covariant
formulation, where the dependence on $\vec{n}$ is present. Thus
LFWFs satisfy all Lorentz symmetries of the front form, including
boost invariance, and they are proper eigenstates of angular
momentum.

In principle, one can solve for the LFWFs directly from the fundamental theory using
methods such as discretized light-front quantization (DLCQ)~\cite{Pauli:1985ps}, the
transverse lattice~\cite{Bardeen:1979xx,Dalley:2004rq,Burkardt:2001jg}, lattice gauge
theory moments~\cite{DelDebbio:1999mq}, Dyson-Schwinger techniques~\cite{Maris:2003vk},
and Bethe--Salpeter techniques~\cite{Brodsky:2003pw}. DLCQ has been remarkably successful
in determining the entire spectrum and corresponding LFWFs in one space-one time field
theories~\cite{Gross:1997mx}, including QCD(1+1)~\cite{Hornbostel:1988fb} and
SQCD(1+1)~\cite{Harada:2004ck}. There are also DLCQ solutions for low sectors of Yukawa
theory in physical space-time dimensions~\cite{Brodsky:2002tp}. The DLCQ boundary
conditions allow a truncation of the Fock space to finite dimensions while retaining the
kinematic boost and Lorentz invariance of light-front quantization.

The transverse lattice method combines DLCQ for one-space and the light-front time
dimensions with lattice theory in transverse space. It has recently provided the first
computation of the generalized parton distributions of the pion~\cite{Dalley:2004rq}.
Dyson-Schwinger methods account well for running quark mass effects, and in principle can
give important hadronic wavefunction information. One can also project known solutions of
the Bethe--Salpeter equation to equal light-front time, thus producing hadronic
light-front Fock wave functions~\cite{Brodsky:2003pw}. Bakker and van Iersel have
developed new methods to find solutions to bound-state light-front equations in ladder
approximation~\cite{vanIersel:2004gf}. Pauli has shown how one can construct an effective
light-front Hamiltonian which acts within the valence Fock state sector
alone~\cite{Pauli:2003tb}. Another possible  method is to construct the $q\bar q$ Green's
function using light-front Hamiltonian theory, DLCQ boundary conditions and
Lippmann-Schwinger resummation.  The zeros of the resulting resolvent projected on states
of specific angular momentum $J_z$ can then generate the meson spectrum and their
light-front Fock wavefunctions.  As emphasized by Weinstein and Vary, new effective
operator methods~\cite{Weinstein:2004nr,Zhan:2004ct} which have been developed for
Hamiltonian theories in  condensed matter and nuclear physics, could also be applied
advantageously to light-front Hamiltonian.  Reviews of nonperturbative light-front
methods may be found in references~\cite{Brodsky:1997de,cdkm,Dalley:ug,Brodsky:2003gk}.

Even without explicit solutions, much is known about the explicit form and structure of
LFWFs.  They can  be matched to nonrelativistic Schrodinger wavefunctions at soft scales.
At high momenta, the LFWFs at large $k_\perp$ and $x_i \to 1$ are constrained by
arguments based on conformal symmetry, the operator product expansion, or perturbative
QCD.  The pattern of higher Fock states with extra gluons is given by ladder
relations~\cite{Antonuccio:1997tw}.  The structure of Fock states with nonzero orbital
angular momentum is also constrained by the Karmanov-Smirnov operator~\cite{ks92}.

\section{AdS/CFT and Its Consequences for Near-Conformal Field Theory}
As shown by Maldacena~\cite{Maldacena:1997re}, there is a remarkable
correspondence between large $N_C$ supergravity theory in a higher
dimensional  anti-de Sitter space and supersymmetric QCD in
4-dimensional space-time.  String/gauge duality provides a framework
for predicting QCD phenomena based on the conformal properties of
the AdS/CFT correspondence. For example, Polchinski and
Strassler~\cite{Polchinski:2001tt} have shown that the power-law
fall-off of hard exclusive hadron-hadron scattering amplitudes at
large momentum transfer can be derived without the use of
perturbation theory by using the scaling properties of the hadronic
interpolating fields in the large-$r$ region of AdS space.  Thus one
can use the Maldacena correspondence to compute the leading
power-law falloff of exclusive processes such as high-energy
fixed-angle scattering of gluonium-gluonium scattering in
supersymmetric QCD.   The resulting predictions for hadron physics
effectively
coincide~\cite{Polchinski:2001tt,Brower:2002er,Andreev:2002aw} with
QCD dimensional counting
rules~\cite{Brodsky:1973kr,Matveev:ra,Brodsky:1974vy,Brodsky:2002st}.
Polchinski and Strassler~\cite{Polchinski:2001tt} have also derived
counting rules for deep inelastic structure functions at $x \to 1$
in agreement with perturbative QCD predictions~\cite{Brodsky:1994kg}
as well as Bloom-Gilman exclusive-inclusive duality.  An interesting
point is that the hard scattering amplitudes which are normally or
order $\alpha_s^p$ in PQCD appear as order $\alpha_s^{p/2}$ in the
supergravity predictions.  This can be understood as an all-orders
resummation of the effective
potential~\cite{Maldacena:1997re,Rey:1998ik}.  The near-conformal
scaling properties of light-front wavefunctions thus lead to a
number of important predictions for QCD which are normally discussed
in the context of perturbation theory.

De Teramond and I~\cite{Brodsky:2003px} have shown how one can use
the scaling properties of the hadronic interpolating operator in the
extended AdS/CFT space-time theory to determine the form of QCD
wavefunctions at large transverse momentum $k^2_\perp \to \infty$
and at $x \to 1$~\cite{Brodsky:2003px}.  The angular momentum
dependence of the light-front wavefunctions also follow from the
conformal properties of the AdS/CFT correspondence.  The scaling and
conformal properties of the correspondence leads to a hard component
of the light-front Fock state wavefunctions of the form:
\begin{eqnarray}
 \psi_{n/h} (x_i, \vec k_{\perp i} , \lambda_i, l_{z i})
&\sim& \frac{(g_s~N_C)^{\frac{1}{2} (n-1)}}{\sqrt {N_C}}
 ~\prod_{i =1}^{n - 1} (k_{i \perp}^\pm)^{\vert l_{z i}\vert}\\[1ex]
&&\times \left[\frac{ \Lambda_o}{ {M}^2 - \sum _i\frac{\vec k_{\perp i}^2 +
m_i^2}{x_i} +
\Lambda_o^2}  \right] ^{n +\sum_i \vert l_{z i} \vert -1}\ ,\nonumber
\label{eq:lfwfR}
\end{eqnarray}
where $g_s$ is the string scale and $\Lambda_o$ represents the basic
QCD mass scale.  The scaling predictions agree with the perturbative
QCD analysis given in the references~\cite{Ji:2003fw}, but the
AdS/CFT analysis is performed at strong coupling without the use of
perturbation theory.  The form of these near-conformal wavefunctions
can be used as an initial ansatz for a variational treatment of the
light-front QCD Hamiltonian.

The recent investigations using the AdS/CFT correspondence has
reawakened interest in the conformal features of
QCD~\cite{Brodsky:2003dn}.  QCD becomes scale free and conformally
symmetric in the analytic limit of zero quark mass and zero $\beta$
function~\cite{Parisi:zy}.  This correspondence principle provides a
new tool, the conformal template, which is very useful for theory
analyses, such as the expansion polynomials for distribution
amplitudes~\cite{Brodsky:1980ny,Brodsky:1984xk,Brodsky:1985ve,Braun:2003rp},
the non-perturbative wavefunctions which control exclusive processes
at leading twist~\cite{Lepage:1979zb,Brodsky:2000dr}.  The
near-conformal behavior of QCD is also the basis for commensurate
scale relations~\cite{Brodsky:1994eh} which relate observables to
each other without renormalization scale or scheme
ambiguities~\cite{Brodsky:2000cr}.  An important example is the
generalized Crewther relation~\cite{Brodsky:1995tb}.  In this method
the effective charges of observables are related to each other in
conformal gauge theory; the effects of the nonzero QCD $\beta-$
function are then taken into account using the BLM
method~\cite{Brodsky:1982gc} to set the scales of the respective
couplings.     The magnitude of the corresponding effective
charge~\cite{Brodsky:1997dh} $\alpha^{\rm exclusive}_s(Q^2) =
{F_\pi(Q^2)/ 4\pi Q^2 F^2_{\gamma \pi^0}(Q^2)}$ for exclusive
amplitudes is connected to the effective charge $\alpha_\tau$ defined from
$\tau$ hadronic decays~\cite{Brodsky:2002nb}
by a commensurate scale relation.  Its magnitude: $\alpha^{\rm
exclusive}_s(Q^2) \sim 0.8$ at small $Q^2,$  is sufficiently large as to
explain the observed magnitude of exclusive amplitudes such as the pion
form factor using the asymptotic distribution
amplitude~\cite{Lepage:1980fj}.

Theoretical~\cite{vonSmekal:1997is,Zwanziger:2003cf,%
Howe:2002rb,Howe:2003mp,Furui:2003mz} and
phenomenological~\cite{Mattingly:ej,Brodsky:2002nb,Baldicchi:2002qm}
evidence is now accumulating that the QCD coupling becomes constant
at small virtuality; {\em i.e.}, $\alpha_s(Q^2)$ develops an
infrared fixed point in contradiction to the usual assumption of
singular growth in the infrared.  If QCD running couplings are
bounded, the integration over the running coupling is finite and
renormalon resummations are  not required.  If the QCD coupling
becomes scale-invariant in the infrared, then elements of conformal
theory~\cite{Braun:2003rp} become relevant even at relatively small
momentum transfers.

Menke, Merino, and Rathsman~\cite{Brodsky:2002nb} and I have
presented a definition of a physical coupling for QCD which has a
direct relation to high precision measurements of the hadronic decay
channels of the $\tau^- \to \nu_\tau {\rm H}^-$.  Let $R_{\tau}$ be
the ratio of the hadronic decay rate to the leptonic one.  Then
$R_{\tau}\equiv R_{\tau}^0\left[1+\frac{\alpha_\tau}{\pi}\right]$,
where $R_{\tau}^0$ is the zeroth order QCD prediction, defines the
effective charge $\alpha_\tau$.  The data for $\tau$ decays is
well-understood channel by channel, thus allowing the calculation of
the hadronic decay rate and the effective charge as a function of
the $\tau$ mass below the physical mass.  The vector and
axial-vector decay modes can be studied separately. Using an
analysis of the $\tau$ data from the OPAL
collaboration~\cite{Ackerstaff:1998yj}, we have found that the
experimental value of the coupling $\alpha_{\tau}(s)=0.621 \pm
0.008$ at $s = m^2_\tau$ corresponds to a value of
$\alpha_{\MSbar}(M^2_Z) = (0.117$-$0.122) \pm 0.002$, where the
range corresponds to three different perturbative methods used in
analyzing the data.  This result is in good agreement with the world
average $\alpha_{\MSbar}(M^2_Z) = 0.117 \pm 0.002$.  However, one
also finds that the effective charge only reaches $\alpha_{\tau}(s)
\sim 0.9 \pm 0.1$ at $s=1\,{\rm GeV}^2$, and it even stays within
the same range down to $s\sim0.5\,{\rm GeV}^2$. The effective
coupling is close to constant at low scales, suggesting that
physical QCD couplings become constant or ``frozen" at low scales.

The near constancy of the effective QCD coupling at small scales
helps explain the empirical success of dimensional counting rules
for the power law fall-off of form factors and fixed angle scaling.
As shown in the references~\cite{Brodsky:1997dh,Melic:2001wb}, one
can calculate the hard scattering amplitude $T_H$ for such
processes~\cite{Lepage:1980fj} without scale ambiguity in terms of
the effective charge $\alpha_\tau$ or $\alpha_R$ using commensurate
scale relations.  The effective coupling is evaluated in the regime
where the coupling is approximately constant, in contrast to the
rapidly varying behavior from powers of $\alpha_{\rm s}$ predicted
by perturbation theory (the universal two-loop coupling).  For
example, the nucleon form factors are proportional at leading order
to two powers of $\alpha_{\rm s}$ evaluated at low scales in
addition to two powers of $1/q^2$; The pion photoproduction
amplitude at fixed angles is proportional at leading order to three
powers of the QCD coupling.  The essential variation from
leading-twist counting-rule behavior then only arises from the
anomalous dimensions of the hadron distribution amplitudes.

\section{Light-Front Phenomenology}

Light-front Fock state wavefunctions $\psi_{n/H}(x_i,\vec k_{\perp
i},\lambda_i)$ play an essential role in QCD  phenomenology,
generalizing Schr\"odinger wavefunctions $\psi_H(\vec k)$ of atomic
physics to relativistic quantum field theory. Given the
$\psi^{(\Lambda)}_{n/H},$ one can construct any spacelike
electromagnetic, electroweak, or gravitational form factor or local
operator product matrix element of a composite or elementary system
from the diagonal overlap of the LFWFs~\cite{BD80}. Exclusive
semi-leptonic $B$-decay amplitudes involving timelike currents such
as $B\rightarrow A \ell \bar{\nu}$ can also be evaluated exactly in
the light-front formalism~\cite{Brodsky:1998hn}.  In this case, the
timelike decay matrix elements require the computation of both the
diagonal matrix element $n \rightarrow n$ where parton number is
conserved and the off-diagonal $n+1\rightarrow n-1$ convolution such
that the current operator annihilates a $q{\bar{q'}}$ pair in the
initial $B$ wavefunction.  This term is a consequence of the fact
that the time-like decay $q^2 = (p_\ell + p_{\bar{\nu}} )^2 > 0$
requires a positive light-cone momentum fraction $q^+ > 0$.
Conversely for space-like currents, one can choose $q^+=0$, as in
the Drell-Yan-West representation of the space-like electromagnetic
form factors. The light-front Fock representation thus provides an
exact formulation of current matrix elements of local operators.  In
contrast, in equal-time Hamiltonian theory, one must evaluate
connected time-ordered diagrams where the gauge particle or graviton
couples to particles associated with vacuum fluctuations.  Thus even
if one knows the equal-time wavefunction for the initial and final
hadron, one cannot determine the current matrix elements.  In the
case of the covariant Bethe-Salpeter formalism, the evaluation of
the matrix element of the current requires the calculation of an
infinite number of irreducible diagram contributions.

One can also prove directly from the LFWF overlap representation
that the anomalous gravitomagnetic moment $B(0)$ vanishes for any
composite system~\cite{Brodsky:2000ii}.  This property follows
directly from the Lorentz boost properties of the light-front Fock
representation and holds separately for each Fock state component.

Given the LFWFs, one can also compute the hadronic distribution
amplitudes $\phi_H(x_i,Q)$  which control hard exclusive processes
as an integral over the transverse momenta of the valence Fock state
LFWFs~\cite{Lepage:1980fj}. In addition one can compute the unintegrated
parton distributions in $x$ and $k_\perp$ which underlie  generalized
parton distributions for nonzero skewness. As shown by Diehl, Hwang,
and myself~\cite{Brodsky:2000xy},  one can give a complete
representation of virtual Compton scattering $\gamma^* p \to \gamma
p$ at large initial photon virtuality $Q^2$ and small momentum
transfer squared $t$ in terms of the light-cone wavefunctions of the
target proton.  One can then verify the identities between the
skewed parton distributions $H(x,\zeta,t)$ and $E(x,\zeta,t)$ which
appear in deeply virtual Compton scattering and the corresponding
integrands of the Dirac and Pauli form factors $F_1(t)$ and $F_2(t)$
and the gravitational form factors $A_{q}(t)$ and $B_{q}(t)$ for
each quark and anti-quark constituent.  We have illustrated the
general formalism for the case of deeply virtual Compton scattering
on the quantum fluctuations of a fermion in quantum electrodynamics
at one loop.

The integrals of the unintegrated parton distributions over
transverse momentum at  zero skewness provide the helicity and
transversity distributions measurable in polarized deep inelastic
experiments \cite{Lepage:1980fj}.  For example, the polarized quark
distributions at resolution $\Lambda$ correspond to
\begin{eqnarray}
q_{\lambda_q/\Lambda_p}(x, \Lambda) &=&  \sum_{n,q_a}
\int\prod^n_{j=1} dx_j d^2 k_{\perp j}\sum_{\lambda_i} \vert
\psi^{(\Lambda)}_{n/H}(x_i,\vec k_{\perp i},\lambda_i)\vert^2
 \\
&& \times\ \delta\left(1- \sum^n_i x_i\right) \delta^{(2)}
\left(\sum^n_i \vec k_{\perp i}\right) \delta(x - x_q)\nonumber \\
&& \times\  \delta_{\lambda_a, \lambda_q} \Theta(\Lambda^2 -
\mathcal{M}^2_n)\ ,\nonumber
\end{eqnarray}
where the sum is over all quarks $q_a$ which match the quantum
numbers, light-cone momentum fraction $x,$ and helicity of the
struck quark.

Hadronization phenomena such as the coalescence mechanism for
leading heavy hadron production are computed from LFWF overlaps.
Diffractive jet production provides another phenomenological window
into the structure of LFWFs.  However, as shown
recently~\cite{Brodsky:2002ue}, some leading-twist phenomena such as the
diffractive component of deep inelastic scattering, single spin
asymmetries, nuclear shadowing and antishadowing cannot be computed from
the LFWFs of hadrons in isolation.

As shown by Raufeisen and myself~\cite{Raufeisen:2004dg}, one can
construct a ``light-front density matrix" from the complete set of
light-front wavefunctions which is a Lorentz scalar. This form can
be used at finite temperature to give a boost invariant formulation
of thermodynamics.  At zero temperature the light-front density
matrix is directly connected to the Green's function for quark
propagation in the hadron as well as deeply virtual Compton
scattering. One can also define a light-front partition function
$Z_{LF}$ as an outer product of light-front wavefunctions. The
deeply virtual Compton amplitude and generalized parton
distributions can then be computed as the trace $Tr[Z_{LF}
\mathcal{O}],$ where $\mathcal{O}$ is the appropriate local
operator~\cite{Raufeisen:2004dg}. This partition function formalism
can be extended to multi-hadronic systems and systems in statistical
equilibrium to provide a Lorentz-invariant description of
relativistic thermodynamics~\cite{Raufeisen:2004dg}.

\section{Complications from Final-State Interactions}

Although it has been more than 35 years since the discovery of
Bjorken scaling~\cite{Bjorken:1968dy} in
electroproduction~\cite{Bloom:1969kc}, there are still many issues
in deep-inelastic lepton scattering and Drell-Yan reactions which
are only now being understood from a fundamental basis in QCD.  In
contrast to the parton model, final-state interactions in deep
inelastic  scattering and initial state interactions in hard
inclusive reactions cannot be neglected---leading to $T-$odd single
spin
asymmetries~\cite{Brodsky:2002cx,Belitsky:2002sm,Collins:2002kn} and
diffractive contributions~\cite{Brodsky:2002ue,Brodsky:2004hi}. This
in turn implies that  the structure functions measured in deep
inelastic scattering are not probability distributions computed from
the square of the LFWFs computed in isolation~\cite{Brodsky:2002ue}.

It is usually assumed---following the parton model---that the
leading-twist structure functions measured in deep  inelastic
lepton-proton scattering are simply the probability distributions
for finding quarks and gluons in the target nucleon.  In fact, gluon
exchange between the fast, outgoing quarks and the target spectators
effects the leading-twist structure functions in a profound way,
leading to  diffractive leptoproduction processes, shadowing of
nuclear structure  functions, and target spin asymmetries. In
particular, the final-state interactions from gluon exchange between
the outgoing quark and the target spectator system lead to
single-spin asymmetries in semi-inclusive deep inelastic
lepton-proton scattering at leading twist in perturbative QCD; {\em
i.e.}, the rescattering corrections of the struck quark with the
target spectators are not power-law suppressed at large photon
virtuality $Q^2$ at fixed $x_{bj}$~\cite{Brodsky:2002cx}  The
final-state interaction from gluon exchange occurring immediately
after the interaction of the current also produces a leading-twist
diffractive component to deep inelastic scattering $\ell p \to
\ell^\prime p^\prime X$ corresponding to color-singlet exchange with
the target system; this in turn produces shadowing and
anti-shadowing of the nuclear structure
functions~\cite{Brodsky:2002ue,Brodsky:1989qz}. In addition, one can
show that the pomeron structure function derived from diffractive
DIS has the same form as the quark contribution of the gluon
structure function~\cite{Brodsky:2004hi}. The final-state
interactions occur at a short light-cone time $\Delta\tau \simeq
1/\nu$ after the virtual photon interacts with the struck quark,
producing a nontrivial phase. Here $\nu = p \cdot q/M$ is the
laboratory energy of the virtual photon. Thus none of the above
phenomena is contained in the target light-front wave functions
computed in isolation. In particular, the shadowing of nuclear
structure functions is due to destructive interference effects from
leading-twist diffraction of the virtual photon, physics not
included in the nuclear light-front wave functions.  Thus the
structure functions measured in deep inelastic lepton scattering are
affected by final-state rescattering, modifying their connection to
light-front probability distributions.  Some of these results can be
understood by augmenting the light-front wave functions with a gauge
link, but with a gauge potential created by an external field
created by the virtual photon $q \bar q$ pair
current~\cite{Belitsky:2002sm}.  The gauge link is also process
dependent~\cite{Collins:2002kn}, so the resulting augmented LFWFs
are not universal.

Single-spin asymmetries in hadronic reactions provide a remarkable
window to QCD mechanisms at the amplitude level.  In general,
single-spin asymmetries measure the correlation of the spin
projection of a hadron with a production or scattering
plane~\cite{Sivers:1990fh}.  Such correlations are odd under time
reversal, and thus they can arise in a time-reversal invariant
theory only when there is a phase difference between different spin
amplitudes.  Specifically, a nonzero correlation of the proton spin
normal to a production plane measures the phase difference between
two amplitudes coupling the proton target with $J^z_p = \pm {1\over
2}$ to the same final-state.  The calculation requires the overlap
of target light-front wavefunctions with different orbital angular
momentum: $\Delta L^z = 1;$ thus a single-spin asymmetry (SSA)
provides a direct measure of orbital angular momentum in the QCD
bound state.

The observation that $\simeq 10\%$ of the positron-proton deep
inelastic cross section at HERA is
diffractive~\cite{Derrick:1993xh,Ahmed:1994nw} points to the
importance of final-state gauge interactions as well as a new
perspective to the nature of the hard pomeron.  The same
interactions are responsible for nuclear shadowing and  Sivers-type
single-spin asymmetries in semi-inclusive deep inelastic scattering
and in Drell-Yan reactions. These new observations are in
contradiction to parton model and light-cone gauge based arguments
that final state interactions can be ignored at leading twist.  The
modifications of the deep inelastic lepton-proton cross section due
to final state interactions are consistent with color-dipole based
scattering models and imply that the traditional identification of
structure functions with the quark probability distributions
computed from the wavefunctions of the target hadron computed in
isolation must be modified.

The shadowing and antishadowing of nuclear structure functions in
the Gribov-Glauber picture is due to the destructive and
constructive coherence, respectively, of amplitudes arising from the
multiple-scattering of quarks in the nucleus.  The effective
quark-nucleon scattering amplitude includes Pomeron and Odderon
contributions from multi-gluon exchange as well as Reggeon quark
exchange contributions~\cite{Brodsky:1989qz}.  The multiscattering
nuclear processes from Pomeron, Odderon and pseudoscalar Reggeon
exchange leads to shadowing and antishadowing of the electromagnetic
nuclear structure functions in agreement with measurements. An
important conclusion is that antishadowing is
nonuniversal---different for quarks and antiquarks and different for
strange quarks versus light quarks.  This picture thus leads to
substantially different nuclear effects for charged and neutral
currents, particularly in anti-neutrino reactions, thus affecting
the extraction of the weak-mixing angle $\sin^2\theta_W$ and the
constant $\rho_o$ which are determined from the ratios of charged
and neutral current contributions in deep inelastic neutrino and
anti-neutrino scattering. In recent work, Schmidt, Yang, and
I~\cite{Brodsky:2004qa} find that a substantial part of the
difference between the standard model prediction and the anomalous
NuTeV result~\cite{Zeller:2001hh} for $\sin^2\theta_W$ could be due
to the different behavior of nuclear antishadowing for charged and
neutral currents.  Detailed measurements of the nuclear dependence
of charged, neutral and electromagnetic DIS processes are needed to
establish the distinctive phenomenology of shadowing and
antishadowing and to make the NuTeV results definitive.

\section{Other QCD Phenomenology Related to Light-Front Wavefunctions}

A number of important phenomenological properties follow directly
from the  structure of light-front wavefunctions in gauge theory.

(1). {\it Intrinsic Glue and Sea.} Even though QCD was motivated by
the successes of the parton model, QCD predicts many new features
which go well beyond the simple three-quark description of the
proton.  Since the number of Fock components cannot be limited in
relativity and quantum mechanics,  the nonperturbative wavefunction
of a proton contains gluons and sea quarks, including heavy quarks
at any resolution scale. Thus there is no scale $Q_0$ in deep
inelastic lepton-proton scattering where the proton can be
approximated by its valence quarks.  Empirical evidence also
continues to accumulate that the strange-antistrange quark
distributions are not symmetric in the
proton~\cite{Brodsky:1996hc,Kretzer:2004bg}.

(2) {\it Intrinsic Charm.}~\cite{Brodsky:1980pb}  The probability
for Fock states of a light hadron such as the proton to have an
extra heavy quark pair decreases as $1/m^2_Q$ in non-Abelian gauge
theory~\cite{Franz:2000ee,Brodsky:1984nx}.  The relevant matrix
element is the cube of the QCD field strength $G^3_{\mu \nu}.$  This
is in contrast to abelian gauge theory where the relevant operator
is $F^4_{\mu \nu}$ and the probability of intrinsic heavy leptons in
QED bound state is suppressed as $1/m^4_\ell.$  The intrinsic Fock
state probability is maximized at minimal off-shellness.  It is
useful to define the transverse mass $m_{\perp i}= \sqrt{k^2_{\perp
i} + m^2_i}.$ The maximum probability then occurs at $x_i = {
m^i_\perp /\sum^n_{j = 1} m^j_\perp}$; {\em i.e.}, when the
constituents have minimal invariant mass and equal rapidity. Thus
the heaviest constituents have the highest momentum fractions and
the highest $x_i$. Intrinsic charm thus predicts that the charm
structure function has support at large $x_{bj}$ in excess of DGLAP
extrapolations~\cite{Brodsky:1980pb}; this is in agreement with the
EMC measurements~\cite{Harris:1995jx}.  It predicts leading charm
hadron production and fast charmonium production in agreement with
measurements~\cite{Anjos:2001jr}. In fact even double $J/\psi's$ are
produced at large $x_F$, consistent with the dissociation and
coalescence of double intrinsic Fock states of the projectile
LFWF~\cite{Vogt:1995tf}.

The proton wavefunction thus contains charm quarks with large
light-cone momentum fractions $x$.  The recent observation by the
SELEX experiment~\cite{Ocherashvili:2004hi,Mattson:2002vu} showing
that doubly-charmed baryons such as the $\Xi_{cc}^+$ and hence two
charmed quarks are produced at large $x_F$ and small $p_T$ in
hadron-nucleus collisions provides additional and compelling
evidence for the diffractive dissociation of complex off-shell Fock
states of the projectile. These observations contradict the
traditional view that sea quarks and gluons are always produced
perturbatively via DGLAP evolution. Intrinsic charm can also explain
the $J/\psi \to \rho \pi$ puzzle~\cite{Brodsky:1997fj}. It also
affects the extraction of suppressed CKM matrix elements in $B$
decays~\cite{Brodsky:2001yt}.

3. {\it Hidden Color.} A rigorous prediction of QCD is the ``hidden
color" of nuclear wavefunctions at short distances. QCD predicts
that nuclear wavefunctions contain ``hidden
color"~\cite{Brodsky:1983vf} components: color configurations not
dual to the usual nucleonic degrees of freedom. In general, the
six-quark wavefunction of a deuteron is a mixture of five different
color-singlet states~\cite{Brodsky:1983vf}.  The dominant color
configuration at large distances corresponds to the usual
proton-neutron bound state where transverse momenta are  of order
${\vec k}^2 \sim 2 M_d \epsilon_{BE}.$ However, at small impact
space separation, all five Fock color-singlet components eventually
acquire equal weight, {\em i.e.}, the deuteron wavefunction evolves
to 80\% hidden color.

\begin{figure}[htb]
\begin{center}
\epsfig{file=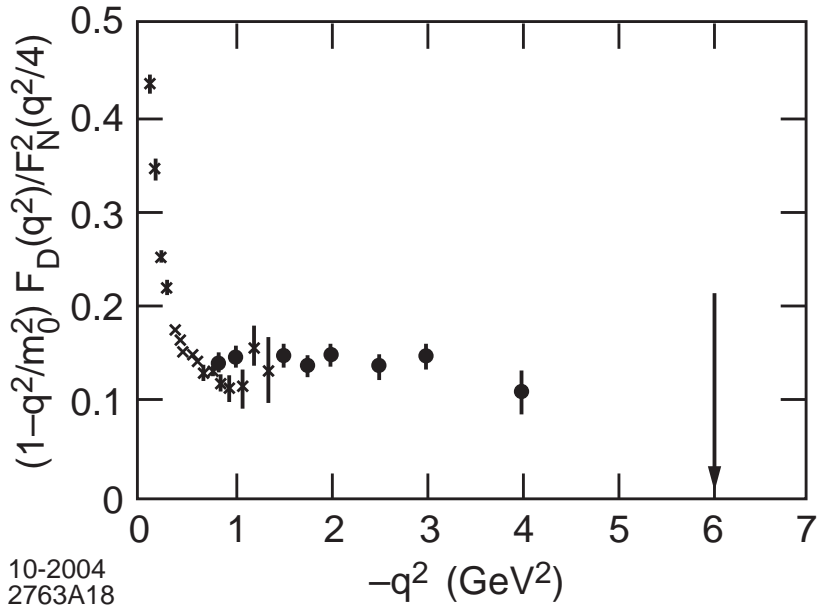,width=4.3in}\end{center}
\caption[*]{Reduced Deuteron Form Factor showing the scaling
predicted by perturbative QCD and conformal scaling.  The data show
two regimes: a fast-falling behavior at small $Q^2$ characteristic
of normal nuclear binding, and a hard scattering regime with
monopole fall-off controlled by the scale $m^2_0 = 0.28~{\rm
GeV^2}.$ The latter contribution is attributable to non-nucleonic
hidden-color components of the deuteron's six-quark Fock state. From
Ref.~\cite{Brodsky:1976rz}. \label{reduced}}
\end{figure}

At high $Q^2$ the deuteron form factor is sensitive to wavefunction
configurations where all six quarks overlap within an impact
separation $b_{\perp i} < \mathcal{O} (1/Q).$ Since the deuteron
form factor contains the probability amplitudes for the proton and
neutron to scatter from $p/2$ to $p/2+q/2$, it is natural to define
the reduced deuteron form factor\cite{Brodsky:1976rz,Brodsky:1983vf}
\begin{equation} f_d(Q^2) \equiv {F_d(Q^2)\over
F_{1N} \left(Q^2\over 4\right)\, F_{1N}\,\left(Q^2\over
4\right)}.\end{equation} The effect of nucleon compositeness is
removed from the reduced form factor. QCD then predicts the scaling
\begin{equation} f_d(Q^2) \sim {1\over Q^2} ; \end{equation}
{\em i.e.}, the same scaling law as a meson form factor.  This
scaling is consistent with experiment for $Q^2 > 1~{\rm GeV}^2.$  In
fact as seen in Fig.~\ref{reduced}, the deuteron reduced form factor
contains two components: (1) a fast-falling component characteristic
of nuclear binding with probability $85\%$, and (2) a hard
contribution falling as a monopole with a scale of order $0.5~{\rm
GeV}$  with probability $15\%.$ The normalization of the deuteron
form factor observed at large $Q^2$~\cite{Arnold:1975dd}, as well as
the presence of two mass scales in the scaling behavior of the
reduced deuteron form factor~\cite{Brodsky:1976rz} thus suggests
sizable hidden-color Fock state contributions such as
$\ket{(uud)_{8_C} (ddu)_{8_C}}$ with probability  of order $15\%$
in the deuteron wavefunction~\cite{Farrar:1991qi}.

(4)  {\it  Color transparency.}  The small transverse size
fluctuations of a hadron wavefunction with a small color dipole
moment will have minimal interactions in a
nucleus~\cite{Bertsch:1981py,Brodsky:1988xz}.

This has been verified in the case of diffractive dissociation  of a
high energy pion into dijets $\pi A \to q \bar q A^\prime$ in which
the nucleus is left in its ground state~\cite{Ashery:2002jx}.  When
the hadronic jets have balancing but high transverse momentum, one
studies the small size fluctuation of the incident pion.  The
diffractive dissociation cross section is found to be proportional
to $A^2$ in agreement with the color transparency prediction. Color
transparency has also been observed in diffractive electroproduction
of $\rho$ mesons~\cite{Borisov:2002rd} and in quasi-elastic $p A \to
p p (A-1)$ scattering~\cite{Aclander:2004zm} where only the small
size fluctuations of the hadron wavefunction enters the hard
exclusive scattering amplitude.  In the latter case an anomaly
occurs at $\sqrt s \simeq 5 $ GeV, most likely signaling a resonance
effect at the charm threshold~\cite{Brodsky:1987xw}.

Color transparency, as evidenced by the Fermilab measurements of
diffractive dijet production, implies that a pion can interact
coherently throughout a nucleus with minimal absorption, in dramatic
contrast to traditional Glauber theory based on a fixed $\sigma_{\pi
n}$ cross section.  Color transparency gives direct validation of
the gauge interactions of QCD.

\section{ Hard Exclusive Processes and  Form Factors at High $Q^2$ }

Leading-twist PQCD predictions for hard exclusive
amplitudes~\cite{Lepage:1980fj}  are written  in a factorized form
as the product of hadron distribution amplitudes $\phi_I(x_i,Q)$ for
each hadron $I$ convoluted with  the hard scattering amplitude $T_H$
obtained by replacing each hadron with collinear on-shell quarks
with light-front momentum fractions $x_i = k^+_i/P^+.$ The hadron
distribution amplitudes are obtained by integrating the $n-$parton
valence light-front wavefunctions: \begin{equation} \phi(x_i,Q) =
\int^Q \Pi^{n-1}_{i=1} d^2 k_{\perp i} ~ \psi_{\rm
val}(x_i,k_\perp).
\end{equation}
Thus the distribution amplitudes are $L_z=0$ projections of the LF
wavefunction, and the sum of the spin projections of the valence
quarks must equal the $J_z$ of the parent hadron.  Higher orbital
angular momentum components lead to power-law suppressed exclusive
amplitudes~\cite{Lepage:1980fj,Ji:2003fw}.  Since quark masses can
be neglected at leading twist in $T_H$, one has quark helicity
conservation, and thus, finally, hadron-helicity conservation: the
sum of initial hadron helicities equals the sum of final helicities.
In particular, since the hadron-helicity violating Pauli form factor
is computed from states with $\Delta L_z = \pm 1,$  PQCD predicts
$F_2(Q^2)/F_1(Q^2) \sim 1/Q^2 $ [modulo logarithms].  A detailed
analysis shows that the asymptotic fall-off takes the form
$F_2(Q^2)/F_1(Q^2) \sim \log^2 Q^2/Q^2$~\cite{Belitsky:2002kj}.  One
can also construct other models~\cite{Brodsky:2003pw} incorporating
the leading-twist perturbative QCD prediction which are consistent
with the JLab polarization transfer data~\cite{Jones:1999rz} for the
ratio of proton Pauli and Dirac form factors.  This analysis can
also be extended to study the spin structure of scattering
amplitudes at large transverse momentum and other processes which
are dependent on the scaling and orbital angular momentum structure
of light-front wavefunctions.  Recently, Afanasev, Carlson, Chen,
Vanderhaeghen, and I~\cite{Chen:2004tw} have shown that the
interfering two-photon exchange contribution to elastic
electron-proton scattering, including inelastic intermediate states,
can account for the discrepancy between Rosenbluth and Jefferson Lab
spin transfer polarization data~\cite{Jones:1999rz}.

A crucial prediction of models for proton form factors is the
relative phase of the timelike form factors, since this can be
measured from the proton single spin symmetries in $e^+ e^- \to p
\bar p$ or $p \bar p \to \ell \bar \ell$~\cite{Brodsky:2003gs}. Carl
Carlson, John Hiller, Dae Sung Hwang and I~\cite{Brodsky:2003gs}
have shown that measurements of the proton's polarization strongly
discriminate between the analytic forms of models which fit the
proton form factors in the spacelike region.  In particular, the
single-spin asymmetry normal to the scattering plane measures the
relative phase difference between the timelike $G_E$ and $G_M$ form
factors.  The dependence on proton polarization in the timelike
region is expected to be large in most models, of the order of
several tens of percent.  The continuation of the spacelike form
factors to the timelike domain $t = s > 4 M^2_p$ is very sensitive
to the analytic form of the form factors; in particular it is very
sensitive to the form of the PQCD predictions including the
corrections to conformal scaling.  The forward-backward $\ell^+
\ell^-$ asymmetry can measure the interference of one-photon and
two-photon contributions to $\bar p p \to \ell^+ \ell^-.$

As discussed in section 2, dimensional counting rules for hard exclusive processes have
now been derived in the context of nonperturbative QCD using the AdS/CFT correspondence.
The data for virtually all measured hard scattering processes appear to be consistent
with the conformal predictions of QCD.  For example, recent measurements of the deuteron
photodisintegration cross section $\gamma d \to p n$ follow the leading-twist $s^{11}$
scaling behavior at large momentum transfers in the few GeV
region~\cite{Holt:1990ze,Bochna:1998ca,Rossi:2004qm}. This adds further evidence for the
dominance of leading-twist quark-gluon subprocesses and the near conformal behavior of
the QCD coupling. As discussed above, the evidence that the running coupling has constant
fixed-point behavior, which together with BLM scale fixing, could help explain the near
conformal scaling behavior of the fixed-CM angle cross sections.  The angular
distribution of hard exclusive processes is generally consistent with quark interchange,
as predicted from large $N_C$ considerations.

\section{New Directions}

As I have emphasized in this talk, the light-front wavefunctions of
hadrons are the central elements of QCD phenomenology, describing bound
states in terms of their fundamental quark and gluon degrees of freedom
at the amplitude level.  Given the light-front wavefunctions one can
compute quark and gluon distributions, distribution amplitudes,
generalized parton distributions, form factors, and matrix elements of
local currents such as semileptonic $B$ decays. The diffractive
dissociation of hadrons on nucleons or nuclei into jets or leading
hadrons provides new measures of the LFWFs of the projectile as well as
tests of color transparency and intrinsic charm.

It is thus imperative to compute the light-front wavefunctions
from first principles in QCD.  Lattice gauge theory can provide moments of
the distribution amplitudes by evaluating  vacuum-to-hadron matrix
elements of local operators~\cite{DelDebbio:1999mq}.  The transverse
lattice is also providing new nonperturbative
information~\cite{Dalley:2004rq,Burkardt:2001jg}.

The DLCQ method is also a first-principles method for solving
nonperturbative QCD; at finite harmonic resolution $K$ the DLCQ
Hamiltonian acts in physical Minkowski space as a finite-dimensional
Hermitian matrix in Fock space.  The DLCQ Heisenberg equation is
Lorentz-frame independent and has the advantage of providing not only the
spectrum of hadrons, but also the complete set of LFWFs for each hadron
eigenstate.

An important feature the light-front formalism is that $J_z$ is conserved; thus one
simplify the DLCQ method by projecting the full Fock space on states with specific
angular momentum. As shown in ref. ~\cite{Brodsky:2003pw}, the Karmanov-Smirnov operator
uniquely specifies the form of the angular dependence of the light-front wavefunctions,
allowing one to transform the light-front Hamiltonian equations to  differential
equations acting on scalar forms. A complementary method would be to construct the
$T$-matrix for asymptotic $q \bar q$ or $qqq$ or gluonium states using the light-front
analog of the Lippmann-Schwinger method.  This allows one to focus on states with the
specific global quantum numbers and spin of a given hadron. The zeros of the resulting
resolvent then provides the hadron spectrum and the respective light-front Fock state
projections.

The AdS/CFT correspondence has now provided important new information on the
short-distance structure of hadronic LFWFs; one obtains conformal constraints which are
not dependent on perturbation theory. The large $k_\perp$ fall-off of the valence LFWFs
is also rigorously determined by consistency with the evolution equations for the hadron
distribution amplitudes~\cite{Lepage:1980fj}. Similarly, one can also use the structure
of the evolution equations to constrain the $x \to 1$ endpoint behavior of the LFWFs. One
can use these strong constraints on the large $k_\perp$ and $x \to 1$ behavior to model
the LFWFs. Such forms can also be used as the initial approximations to the wavefunctions
needed for variational methods which minimize the expectation value of the light-front
Hamiltonian.

\section*{Acknowledgments}

I wish to thank Professors Ben Bakker, Piet Mulders, and their colleagues at the Vrije
Universiteit in Amsterdam for  hosting this outstanding meeting. This talk is based on
collaborations with Guy de Teramond, Markus Diehl, Rikard Enberg, John Hiller, Paul
Hoyer, Dae Sung Hwang, Gunnar Ingelman, Volodya Karmanov, Gary McCartor, Sven Menke,
Carlos Merino, Joerg Raufeisen, and Johan Rathsman.

\end {document}